# Gate-Tuned Thermoelectric Power in Black Phosphorus

Yu Saito[1]*†, Takahiko Iizuka[1]†, Takashi Koretsune[2], Ryotaro Arita[2]
Sunao Shimizu[2] and Yoshihiro Iwasa[1,2]

[1] *Quantum-Phase Electronics Center (QPEC) and Department of Applied Physics,*
*The University of Tokyo, Tokyo 113-8656, Japan.*
[2] *RIKEN Center for Emergent Matter Science (CEMS), Wako 351-0198, Japan*

†These authors equally contributed to this work.
*Address correspondence to saito@mp.t.u-tokyo.ac.jp

**ABSTRACT**

The electric field effect is a useful means of elucidating intrinsic material properties as well as for designing functional devices. The electric-double-layer transistor (EDLT) enables the control of carrier density in a wide range, which is recently proved to be an effective tool for the investigation of thermoelectric properties. Here, we report the gate-tuning of thermoelectric power in a black phosphorus (BP) single crystal flake with the thickness of 40 nm. Using an EDLT configuration, we successfully control the thermoelectric power ($S$), and find that the $S$ of ion-gated BP reached +510 μV/K at 210 K in the hole depleted state, which is much higher than the reported bulk single crystal value of +340 μV/K at 300 K. We compared this experimental data with the first-principles-based calculation and found that this enhancement is qualitatively explained by the effective thinning of the conduction channel of the BP flake and non-uniformity of the channel owing to the gate operation in a depletion mode. Our results provide new opportunities for further engineering BP as a thermoelectric material in nanoscale.

Black phosphorus (BP), a layered elemental semiconducting material with a direct band gap of ~0.3 eV, has received renewed attention due to its high field effect mobility[1–3] as well as its photonic[4–8] and ambipolar transport properties.[9,10] A BP single crystal possesses a puckered honeycomb structure, where each phosphorus atom is covalently bonded with three adjacent phosphorus atoms.[11–13] Owing to the relatively weak van der Waals force between its layers, nanometer-thick BP crystals can be cleaved via mechanical exfoliation. As well as being a potentially functional phenomenon, thermoelectric properties are useful for developing a basic understanding of carrier transport properties in materials with reduced dimensionality. Indeed, BP thin films or its monolayer—phosphorene—have been theoretically predicted to be promising thermoelectric materials.[14–20] It has been reported that the thermoelectric power, expressed as the Seebeck coefficient ($S$), of few-layer BP thin films can increase by up to approximately 2000 µV/K. However, there are few experimental reports of the thermoelectric properties of BP thin flakes, especially their carrier density dependence, which stands in contrast to the depth of work on BP bulk crystals.[20–23] Therefore, experimental studies of the thermoelectric properties of BP nano-thick flakes are highly desirable, particularly those that investigate carrier density dependence.

The external field-effect modulation of materials' carrier density is a powerful way to understand its electrical and thermal transport behavior. The use of field-effect modulation minimizes the effects of unintentionally introduced disorder and structural modifications originating from chemical dopants, which are known to occur in the case of chemical doping. The recent development of electric-double-layer transistor (EDLT) allows us to apply a large electric field of over 10 MV/cm at a liquid/solid interface,[25] thereby resulting in sharp band bending and the formation of a high-density two-dimensional electron system with the thickness of 1-2 nm[26,27] at the surface of materials as well as the depletion of a large amount of carriers originally doped in the bulk. Although anions or cations in ionic liquids are physically close to the channel, and may introduce scattering or local charge density fluctuations, the mobility of ion-gated transistors is still comparable with that of conventional solid-gated transistor.[10] This suggests that the EDLT can provide an ideal platform to investigate carrier density dependent physical properties such as superconductivity[28], and indeed it has already been used in the thermoelectric measurements of ZnO, carbon-nanotubes, InAs nanowires, $SrTiO_3$ and few-layer $WSe_2$.[29–34] Here, we report experimental and theoretical investigations of the thermoelectric power of a BP single crystal by using an EDLT configuration. We find that the $S$ of ion-gated BP reached 510 µV/K in a deplete state at 210 K, which exceeds the $S$ of bulk BP. By comparison with the first-principles calculated values, this enhancement of thermoelectric power is potentially explained by the effective thinning of the conduction layer to the nanoscale and the non-uniformity of channel due to the transistor operating in a depletion mode.

We prepared BP flakes from a bulk single crystal (purchased from Smart-elements, Austria) through a mechanical exfoliation method and then transferred the flakes to a $SiO_2$/Si substrate. To investigate the thermoelectric properties of BP, we chose a 40-nm-thick flake (Fig. 1) as determined by atomic force microscopy and then fabricated electrodes (Cr/Au, 7/90 nm) in a Hall bar configuration with covers (Cr/$SiO_2$, 5nm/20nm) via electron beam lithography, deposition and lift-off. The cover layer was made to avoid possible electrochemical damage of the gold electrodes by the ionic liquid. In this study, we did not take the in-plane anisotropy into account in fabricating devices because the anisotropy in $S$ is predicted to be small by the first-principles band calculation, as described below. A droplet of ionic liquid (N,N-diethyl-N-(2-methoxyethyl)-N- methylammonium bis (trifluoromethylsulphonyl) imide (DEME-TFSI)) covered both the channel area of the BP thin flake and the gate electrode. A gate voltage ($V_G$) was applied at $T = 210$ K, which is just above the glass transition temperature of DEME-TFSI (~190 K). The BP transport measurements were performed in a helium-free refrigerator combined with an AC resistance-bridge (Lake Shore Cryotronics Inc. model 730). For the thermoelectric measurement setup, we followed well-established methodology used in previous studies of the thermoelectric properties of micro-scale devices[33,35–40] in which source and drain electrodes also served as thermometers. A temperature gradient, $\Delta T$, is generated by fabricating micro-heater probes as shown Fig. 1. A current is applied to the heater causing a temperature gradient to form across the substrate via Joule heating. The thermoelectric power was measured with voltage probes. In this measurement set up, all the electrodes including the heater and the thermometers covered by 20-nm-thick $SiO_2$ films as mentioned above are electrically disconnected to the ionic liquid, and thus the voltage leakage from heater to thermometer via ionic liquid is highly unlikely.

Figures 2a and b show the transport characteristics of the BP-EDLT: the excitation source-drain current ($I_{DS}$) (top) and a 4-probe resistance ($R_{xx}$) (bottom) as a function of $V_G$ between −1 and 5 V, with a sweep speed of 10 mV/s at 210 K, which is slightly above the glass transition temperature of the DEME-TFSI. In these measurements $I_{DS}$ was measured with an excitation source-drain voltage ($V_{DS}$) fixed at 10 mV. The BP-EDLT exhibited reversible *p*-type transistor behavior without any sign of degradation or anomalies originating from electrochemical reactions. At $V_G = 0$ V, the BP-EDLT was in the normally ON-state, which is due to the large amount of hole carriers unintentionally doped into the BP single crystal. While a negative $V_G$ further accumulates hole carriers, a positive $V_G$ causes a depletion of bulk hole carriers to the opposite direction from the surface, leading to an insulating state (OFF-state). The observed hysteresis behavior is seemingly due to the slow motion of ions in the ionic liquid. This particular sample shows ambipolar behavior but does not become conducting for the electron side, even after the 1 hour annealing process under vacuum at $1 \times 10^{-5}$ torr.[10]

In the thermoelectric measurements, first of all, we calibrated the thermometers and confirmed the linear relation between the temperature gradient $\Delta T$ and the square of heater voltage $V_{Heater}^2$ (see Figures S1-S3 in the Supporting Information), following the previous research.[33] In extracting the Seebeck coefficients, we excluded the sets of data points whose linear fit was less than 90% accurate (Fig. S3), and thus neither the sign change of the $S$ at $V_G$~2.5 V nor reliable values for electron side were not observed because of the high contact resistance. Figures 3a–c show the volumetric conductivity ($\sigma_{3D}$), $S$ and thermoelectric power factor ($S^2\sigma_{3D}$) as a function of $V_G$ between −1 and 2.5 V, respectively. We also investigated the relation between mobility and $S$ (Fig. S4). Here, for calculating conductivity we assumed that the effective thickness of conducting layer (hole accumulated region) does not change by the application of $V_G$, and that the hole depletion occurs uniformly in the whole flake. At $V_G$ = 0 V, the $S$ was +220 μV/K, which is a reasonable value when we consider that the result of a bulk single crystal at 300 K was reported to be +340 μV/K,[24] especially considering that these researchers observed that $S$ increased with increasing temperature and that our measurements were made at 210 K. The conductivity monotonically decreases with increasing positive $V_G$, while the $S$ increases. The $S$ reached a maximum value of +510 μV/K at 2.5 V, which is more than two times larger than that at 0 V. The $S^2\sigma_{3D}$, however, exhibits its maximum value of 460 µW/K²m at about 0.7 V. This value is much smaller than the theoretical prediction[17] because the direction of the channel in the present system is not optimized, thereby leading to lower conductivity. From the previous research, thermal conductivity $\kappa$ can be estimated to be, at most, ~15 W/Km around 200 K in films less than 50 nm in thickness,[41] and thus the thermoelectric figure of merit ($ZT$) would be obtained as $ZT \sim$ 0.06 at 210 K.

To understand the behavior of $S$ vs. $V_G$, we compared the experimental data with theoretical calculations for bulk BP as a function of carrier density. We first calculated the theoretical values of the $S$ of bulk BP at $T$ = 210 K. Here, we performed a first-principles calculation within the generalized-gradient approximations[42] based on density functional theory (DFT) using the quantum-ESPRESSO package.[43] An ultrasoft pseudopotential[44] and plane wave basis sets with cutoff energies of 50 Ry for wave functions and 500 Ry for charge densities were used. The lattice parameters and atomic geometries were fully relaxed. For the transport properties, we employed the Boltzmann transport theory with a constant relaxation-time approximation using the DFT electronic structure.[45] The band energies of the conduction band were shifted by 0.2 eV to reproduce the experimental band gap.

Figure 3d summarizes the theoretical and experimental value of $S$ at 210 K as a function of volumetric carrier density ($n_{3D}$) from $10^{17}$ to $10^{21}$ cm$^{-3}$. Note that the $x$ and $y$ components of the Seebeck coefficients, $S_x$ and $S_y$, reflecting the in-plane structural anisotropy in BP, are almost identical, indicating the isotropic nature of the Seebeck effect in contrast to the fairly anisotropic conductivity.[2] When plotting the experimental data of $S$ vs. $n_{3D}$, we need to assume the thickness of the conducting layers. We estimated the sheet carrier density $n_{2D}$ from the equation of $n_{2D} = C_{EDL}V_G/e$, where $e$ is the elementary charge and $C_{EDL}$ is the EDL capacitance, which was measured to be 4.8 µF/cm² by Hall effect as previously reported.[10] To convert the $n_{2D}$ to $n_{3D}$, the thickness of the layer must be known. For a first order estimation, we assumed the carriers are depleted uniformly throughout the whole flake by the application of $V_G$ and obtained $n_{3D}$ as $n_{2D}/d$, where $d$ = 40 nm is the thickness of the BP flake used. As shown in Fig. 3d, the experimental $S$ value at $n_{3D}$ = 2.3×10¹⁹ cm⁻² at $V_G$ = 0 V is +220 µV/K, which is slightly above the theoretical estimation and increases more rapidly and then reaches as high as +510 µV/K when the $n_{3D}$ is decreased to 7.5×10¹⁸ cm⁻³. This $S$ value is unexpectedly larger than the theoretical estimation based on Mott relation. The theoretical calculation predicts that such a large $S$ value can only be obtained when the carrier density is as low as 10¹⁷ cm⁻³, but such a low carrier density is unlikely in the present experiment. This indicates that the naïve assumption of the uniform carrier depletion over the whole flake should be reconsidered.

In the present case of the normally ON state of BP, a more realistic model of the EDLT operation is that the depletion layer is widened by the gate voltage, as schematized in Fig. 3e. The thickness of conduction layer is thinned by the application of the positive $V_G$, and eventually the current flow is significantly reduced. The possible interpretation of the anomalous increase of the $S$ demonstrated in Fig. 3d is thus the effective reduction of the conducting layer, keeping the $n_{3D}$ unchanged (Fig. 3e). If this is the case, the conductivity reduction by a factor of nearly 100 at $V_G$ = 2.5 V compared with that at $V_G$ = 0 V indicates that the effective thickness of the conducting layer in the highly depleted state is reduced to a few monolayer or thinner. In fact, a naïve estimation of the depletion layer thickness ($W$) using the general formula,[4]

$$W = \frac{\epsilon_{BP}}{C_{EDL}} \left( \sqrt{1 + \frac{2\epsilon_{IL}^2 V_G}{eN_A \epsilon_{BP} d^2}} - 1 \right) \quad (1)$$

gave ~ 15 nm at 2.5 V. Here, $\epsilon_{BP}$, $\epsilon_{IL}$, $N_A$, $d$ are the dielectric constant of BP, the dielectric constant of ionic liquid, the acceptor concentration, the width of EDL, respectively. This value of $W$ is in the same order as the flake thickness, and consistent with the theoretical estimation based on first-principles-based calculation, which reports that depletion layer can be over 30 nm at the surface bad vending of 2.0 eV[8]. Although the quantitative accuracy remains to be reached, the effective reduction of conducting layer thickness could be realistic.

In the depletion regime, if the Fermi energy is near the valence band edge, the likely mechanism of the carrier transport is the thermal activation or the variable range hopping (VRH). Indeed, a very recently published work on thermoelectric properties of gated BP thin flakes adopted the latter model and explained the temperature dependence of conductivity and Seebeck coefficient in terms of the 2D VRH model[47]. The authors suggested that the thermoelectric properties are governed by the trapped charges due to substrate roughness. This approach may be valid for the depletion regime in a semiconductor. However, in the depletion mode of field effect transistor, we should consider that the channel should be inhomogeneous along to the film thickness direction as explained in the previous paragraph and Fig. 3e. Also, within the activation or VRH model, quantitative understanding of the absolute $S$ value is difficult. Hence, we introduce, in the following, a nonuniform conducting channel model in the depletion mode, and discuss the Mott relation–based model in order to make a semi-quantitative comparison between the experimental data and the first-principles band calculation.

Figure 4a and b show the density of states spectra for bulk, five-, bi-, and monolayer BP, and carrier density dependence of $S_x$ and $S_y$ for mono-, bi- and five-layer and bulk BP (from top to bottom), respectively. The dispersive peak seen at the band edge is ascribed to the one-dimensional-like electronic structure of the valence band top of BP. We found that the $S$ has little anisotropy and increases with decreasing thickness because the density of states becomes more two-dimensional due to the BP's dimensionality approaching the nanoscale. Importantly, $S$ values for the monolayer are as large as +400 μV/K even at a $n_{3D}$ of ~ $1 \times 10^{19}$ cm$^{-3}$ and further increase with decreasing carrier density. These results suggest that the anomalously enhanced $S$ value in BP observed in the present experiment is qualitatively understood in terms of a combined effect of the effective reduction of the layer thickness by the field effect and the reduction in carrier density.

In reality, the depleted state is more complex and the carrier density continuously changes along with the thickness direction and therefore the total Seebeck coefficient ($S_{total}$) will be given by the following equation:

$$S_{\text{total}} = \frac{\int \sigma_{2D}(z) S(z) dz}{\int \sigma_{2D}(z) dz} \qquad (2)$$

This equation suggests that at the lowest carrier region, $S$ will be further enhanced in a few or monolayer BPs beyond the actually observed value of +510 μV/K for a depletion state of a transistor fabricated from a 40-nm-thick BP crystal. For more quantitative analysis of the $V_G$ dependence of the $S$, the Poisson-Schrödinger type analysis combined with the Mott-expression of the Seebeck coefficient is required.

In summary, we have demonstrated gate-tunable thermoelectric power in a BP single crystal using an EDLT configuration. The $S$ value reached +510 μV/K in the hole-deplete state, which is much larger than the value for bulk single crystals. We have made a semi-quantitative argument on the enhanced $S$ on the basis of the Mott relation of the Seebeck effect using the first-principles-based band calculations. In particular, we pointed out the effective reduction of channel thickness in the depletion mode operation of the field effect transistor. Our consideration suggests that a few or monolayer BP single crystal might be capable of producing even larger thermoelectric power. Very recently, Choi et al. reported that the hopping transport is the dominating factor for the $S$ value. [47] Further detailed experimental and theoretical attempts are required to understand the intrinsic thermoelectric properties of BP.

**Associated Content**

**Supporting Information**: Details of thermoelectric measurements and relation between mobility and Seebeck coefficient of the BP-EDLT.

**Author Information**
**Corresponding Author**

*E-mail: saito@mp.t.u-tokyo.ac.jp

**Acknowledgements**

We thank M. Nakano and M. Yoshida for fruitful discussions. Y.S. was supported by the Japan Society for the Promotion of Science (JSPS) through a research fellowship for young scientists. This work was supported by Grant-in-Aid for Specially Promoted Research (no. 25000003) from JSPS and Grant-in-Aid for Young Scientists (B) (No. 26820298).

**Author Contributions**

Y.S., T.I. and Y.I. conceived the idea and designed the experiments. Y.S. fabricated BP-EDLT devices. T.I. performed thermoelectric measurements. Y.S. and T.I. analyzed the data. T.K. carried out theoretical calculations. R.A. and S.S. led physical discussions. Y.S., T.K. and Y.I. wrote the manuscript.

**Author Contributions**

Y.S. and T.I. equally contributed to this work.

**Notes**

The authors declare no competing financial interests.

**Figure Captions**

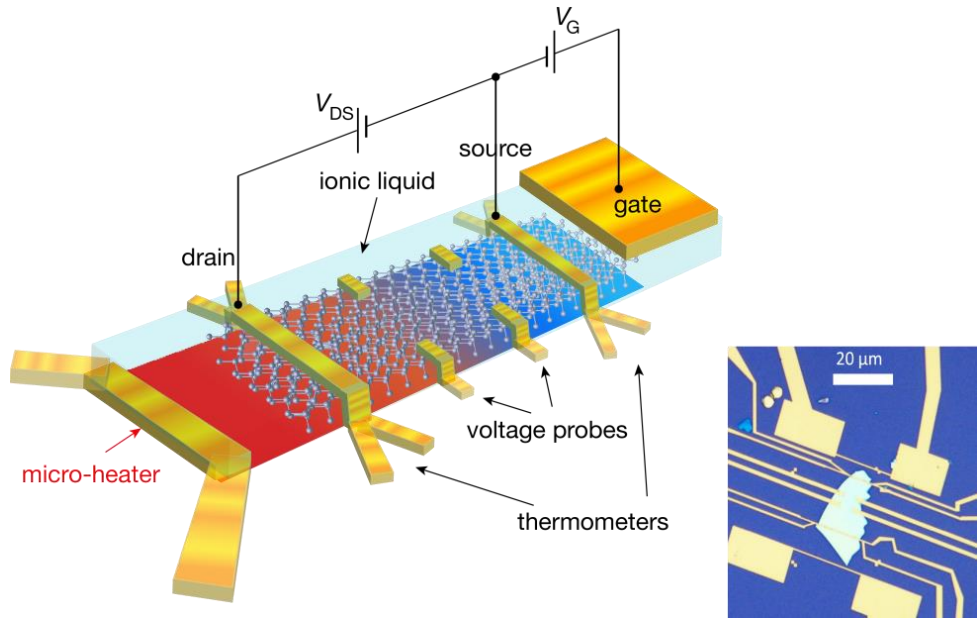

**Figure 1. Conceptual image of a BP-EDLT for thermoelectric measurements.** A schematic image of a BP-EDLT for thermoelectric measurements and the optical microscope image of the real thermoelectric device of a BP flake with a thickness of 40 nm.

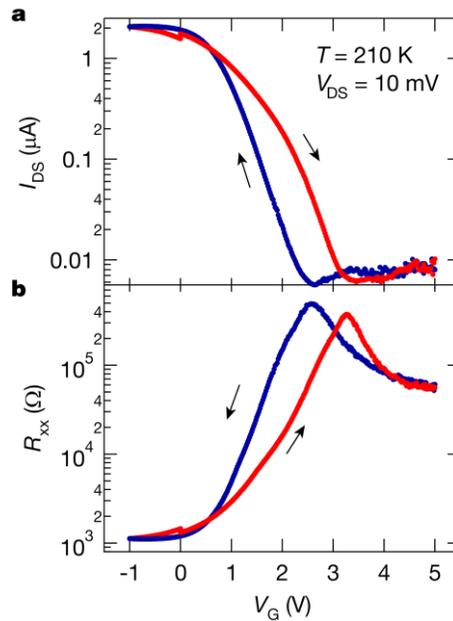

**Figure 2. Transfer curve of a BP-EDLT.** Source-drain current (**a**) and sheet resistance (**b**) as a function of gate voltage ($V_G$) at $T = 210$ K and $V_{DS} = 10$ mV.

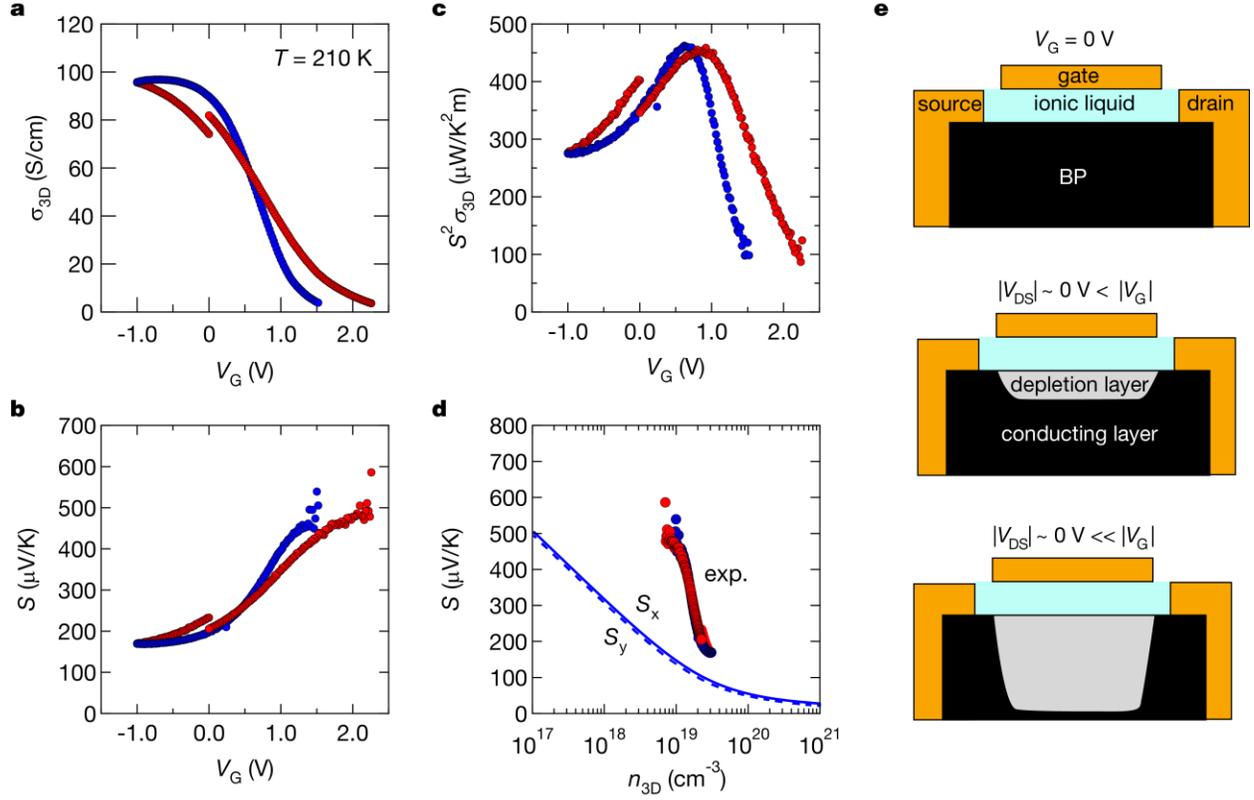

**Figure 3. Tunable thermoelectric power of a BP-EDLT. a-c**, Three dimensional conductivity ($\sigma_{3D}$) (**a**), thermoelectric power (**b**) and power factor (**c**) as a function of gate voltage at $T = 210$ K. **b**, as a function of $V_G$ at $T = 210$ K. **d**, Comparison between experimental data and bulk values of theoretical calculation ($S_x$: solid line, $S_y$: dashed line) of the Seebeck coefficient. In the estimation of $\sigma_{3D}$ in **a**, we assumed uniform carrier distribution in the whole flake. **e**, Schematics of the device operation of BP-EDLT. By application of the gate-voltage, the depletion layer is widened associated with the thinning of the conducting layer.

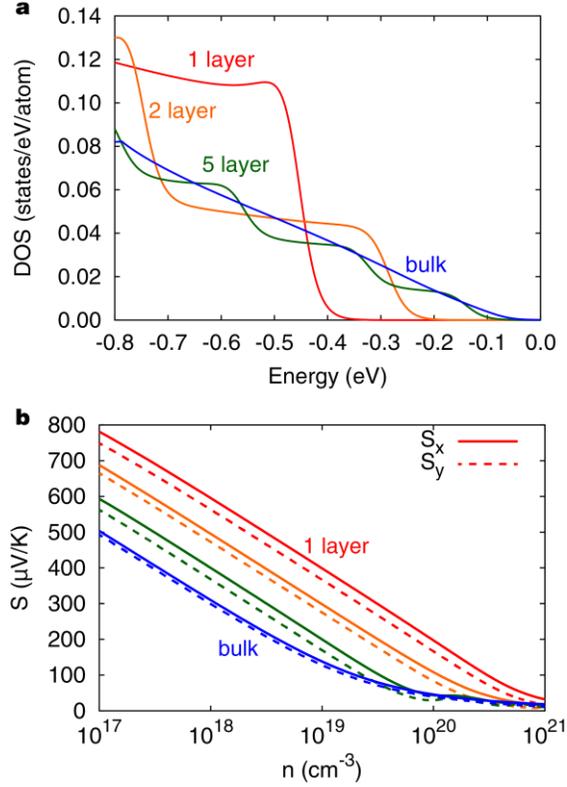

**Figure 4. Theoretical calculations of thermoelectric properties of BP. a,** Density of states at the valence top for single-layer, bilayer, five-layer phosphorene and bulk BP. The energy is measured from the middle of the band gap. The divergent peak seen at each band edge is due to the highly anisotropic band structure. Since the anisotropy is very large at the band edge, the electronic structure can be regarded as one-dimensional. **b,** Carrier density dependence of $S_x$ and $S_y$ for single-layer, bilayer and five-layer phosphorene and bulk BP (from top to bottom) at $T = 210$ K.